\documentclass[prl,showpacs,twocolumn]{revtex4}

\usepackage{epsfig}

\begin{document}
\bibliographystyle{revtex}

\title{Persistence of coherent quantum dynamics at strong dissipation}

\author{Denis Kast and Joachim Ankerhold}
\affiliation{Institut f\"ur Theoretische Physik, Universit\"at Ulm, Albert-Einstein-Allee 11, 89069 Ulm, Germany}

\date{\today}

\begin{abstract}
The quantum dynamics of a two state system coupled to a bosonic reservoir with sub-Ohmic spectral density is investigated for strong friction. Numerically exact path integral Monte Carlo methods reveal that a changeover from coherent to incoherent relaxation does not occur for a broad class of spectral distributions. In non-equilibrium coherences associated with substantial system-reservoir entanglement exist even  when strong dissipation forces the thermodynamic state of the system to behave almost classical.  This may be of relevance for current experiments with nanoscale devices.
\end{abstract}

\pacs{03.65.Yz, 03.67.-a, 73.63.-b, 05.10.Ln}

\maketitle
{\em Introduction--}The impact of dissipative environments on the dynamics
of quantum systems has regained interest due to the boost of activities to tailor
atomic, molecular and solid state structures with growing complexity \cite{ions,saclay,hulst}. Common wisdom is that
quantum coherence is inevitably destroyed at sufficiently strong coupling to broadband reservoirs.
 A paradigmatic model is a two-state system interacting with a reservoir of bosonic degrees of freedom (spin-boson model) which plays a fundamental role in a variety of applications \cite{leggett,weiss,breuer}. At low temperatures and weak coupling, an initial non-equilibrium state evolves via damped coherent oscillations, while at stronger dissipation it displays an incoherent decay towards thermal equilibrium. This is often understood as a quantum to classical changeover. The question we pose here is whether this picture always applies.

The spin-boson (SB) model has recently gained renewed attention for reservoirs with sub-Ohmic mode distributions $J_s(\omega)\propto \alpha\, \omega^s$ where $\alpha$ denotes a coupling constant and $0< s<1$. This class of reservoirs constitutes a dominant noise source in solid state devices at low temperatures such as superconducting qubits \cite{fnoise} and quantum dots \cite{ring} with the spectral exponent $s$ determined by
the microscopic nature of environmental degrees of freedom. It also appears in the context of ultra slow glass dynamics \cite{nalbach}, quantum impurity systems \cite{heavyf}, and nanomechanical oscillators \cite{nanomech}. Advanced numerical techniques \cite{anders,rieger,hofstetter,exdiag} have revealed that at zero temperature the equilibrium state exhibits at a critical coupling strength $\alpha_c$ a quantum phase transition (QPT) from a delocalized phase with tunneling between the two spin orientations (weak friction) to a localized one with almost classical behavior (stronger friction)\cite{chin}.

The time evolution of the polarization towards these asymptotic phases shows coherent oscillations or classical-like monotone decay. It has been argued that with increasing friction the relaxation dynamics {\em always} turns from coherent to incoherent \cite{wang}. Indeed, numerical studies \cite{thorwart} confirmed this picture for reservoirs with $1/2\leq s<1$, but the situation for $0<s<1/2$ remained unclear, mainly because approaches used previously are restricted to the regime of weak to moderate dissipation \cite{anders,thoss2}. The goal of this Letter is to attack this latter regime via real-time path integral Monte Carlo techniques (PIMC) \cite{egger,lothar1} which cover also strong friction. We verify that coherences persist in non-equilibrium for arbitrary coupling strength to a heat bath even when the thermal state is essentially classical. Our results shed new light on our understanding of the quantum-classical crossover and may be accessible experimentally.

{\em Reduced dynamics -- }We consider a  symmetric SB model
\begin{equation}\label{hamilton}
H_{SB}= -\frac{\hbar\Delta}{2}\sigma_x -\frac{1}{2}\, {\sigma_z}\, {\cal E}
+
\sum_\alpha
\hbar\omega_\alpha \,b^\dagger_\alpha b_\alpha
\end{equation}
with a two state system (TSS) which  interacts bilinearly with a harmonic reservoir $H_B$
via the bath force ${\cal E}=\sum_\alpha \lambda_\alpha (b_\alpha+b^\dagger_\alpha)$.
All relevant observables are obtained from the reduced density operator
$\rho(t)={\rm Tr}_B\{\exp(-iH_{SB}t/\hbar) W(0) \exp(iH_{SB}t/\hbar)\}$, where
 the initial state has the form
\begin{equation} \label{eq:initial_prep}
W(0) =  \rho_S(0)\, {\rm e}^{-\beta  (H_B-{\cal E})}/Z_B\, .
\end{equation}
Here, according to typical experimental situations \cite{weiss} the  bath distribution is equilibrated to the initial state of the TSS which is $\rho_S(0)=\left|+1\right\rangle\langle +1|$ (the eigenstates of $\sigma_z$ are $|\pm 1\rangle$ with the bare tunneling amplitude $\Delta$ between them); the partition function is denoted by $Z_B$.
We are interested in the real-time dynamics of the observables
\[
P_\nu(t)\equiv\langle \sigma_\nu(t)\rangle={\rm Tr}\left\{\sigma_\nu\, \rho(t)\right\}\ , \ \nu=x, y, z\, ,
\]
where $P_z$ describes the population difference (polarization) and $P_x$ the coherence between the sites $|\pm 1\rangle$.

A non-perturbative treatment is obtained within the path integral formulation.
The $P_\nu$ is expressed along a Keldysh contour with forward
$\sigma$ and backward $\sigma'$ paths \cite{weiss}. The impact of the environment appears as an
influence functional introducing arbitrarily long-ranged
interactions between the paths. Switching to
${\eta}/\xi=\sigma\pm\sigma'$ one
arrives at
\begin{equation}\label{pathinte}
P_\nu(t)=\int {\cal{D}}[\eta]{\cal{D}}[\xi]{\cal{A}}_\nu\, {\rm e}^{-\Phi[\eta,\xi]}
\end{equation}
with the contribution $\cal{A}_\nu$ in absence of dissipation and the influence functional
\[
\Phi[\eta,\xi]=\int_0^t\hspace{-0.2cm}du\hspace{-0.1cm}\int_0^u\hspace{-0.2cm}dv \dot{\xi}(u)[Q'(u-v) \dot{\xi}(v)+i Q''(u-v) \dot{\eta}(v)]\, .
\]
 According to (\ref{eq:initial_prep}) one sums over all paths with $|\eta(0)|=1, \xi(0)=0$ and $\eta(t)=1, \xi(t)=0$ for $P_z$ and $\eta(t)=0, |\xi(t)|=1$ for $P_x$, respectively. The kernel $Q=Q'+ i Q''$ is related to the bath correlation $\ddot{Q}(t)=\langle{\cal E}(t){\cal E}(0)\rangle/\hbar^2$ and is thus completely determined by the spectral function  $J(\omega) = \pi \sum_\alpha \lambda_\alpha^2
\delta(\omega-\omega_\alpha)$. For sub-Ohmic spectral distributions
\begin{equation}\label{spectral}
J_s(\omega)= 2\pi \alpha \omega_c^{1-s} \omega^s {\rm e}^{-\omega/\omega_c}\ , \ 0< s< 1\, ,
\end{equation}
one has at zero temperature \cite{weiss}
\[
Q_0(t)=2\alpha \Gamma(s-1) [1-(1+i\omega_c t)^{1-s}]\, .
\]
In the Ohmic case $s\to 1$ the coupling $\alpha$ coincides with the Kondo parameter $K$. The
cut-off $\omega_c$ corresponds e.g. to a Debye frequency \cite{weiss} or to a parameter of an electromagnetic environment \cite{ring}.

A direct evaluation of (\ref{pathinte}) is extremely challenging due to the retardation in the influence functional which grows with decreasing temperature. In this situation PIMC methods are very powerful means to explore the non-perturbative range including strong coupling \cite{egger,lothar1}. Here, we extend this formulation
 to simulate not only populations $P_z$ but also coherences $P_x$ (see \cite{epaps}).

\begin{figure}
\epsfig{file=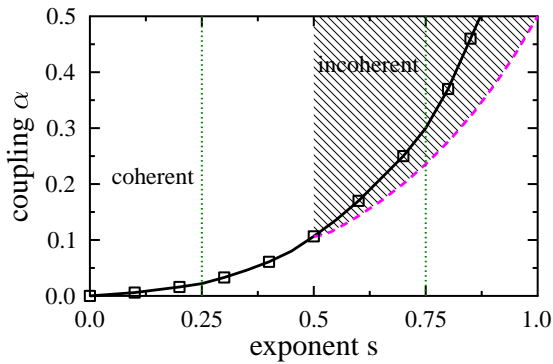, angle=0, width=7.5cm}
\caption{Domains of coherent (white) and incoherent (shaded) dynamics of the SB model for a sub-Ohmic environment (\ref{spectral}) at zero temperature. Above (below) the solid line $\alpha_c(s)$ the system asymptotically reaches a thermal equilibrium which is localized (delocalized) \cite{data}. A coherent-incoherent changeover only occurs along the dashed line $\alpha_{CI}(s)$ (\ref{cialpha}). Dotted lines refer to results in Figs. \ref{slarge}, \ref{ssmall}, respectively.}\label{phasediagram}
\end{figure}
\begin{figure}
\vspace{5.cm}
\epsfig{file=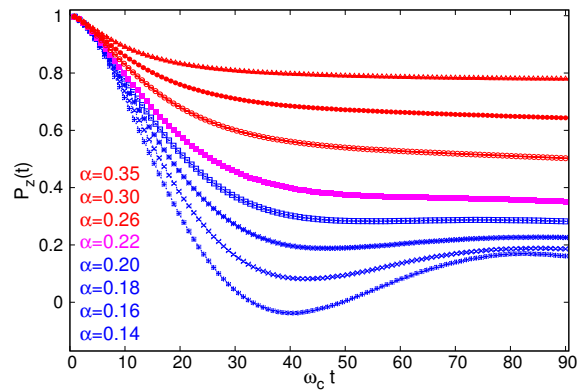, angle=-90, width=7.5cm}
\caption{Dynamics of $P_z(t)$ at $T=0$ and $\Delta/\omega_c=0.1$ according to PIMC for $s=0.75$ and couplings $\alpha$ with $\alpha_c\approx 0.3>\alpha_{CI}\approx 0.22$ (purple line). Blue (red) lines refer to coherent $\alpha<\alpha_{CI}$ (incoherent $\alpha>\alpha_{CI}$) dynamics (cf. Fig.~\ref{phasediagram}). Statistical errors are of the size of the symbols.}\label{slarge}
\end{figure}

{\em Population dynamics-- }The dynamics of $P_z(t)$ directly displays the impact of decoherence as the bare two state system is completely quantum and has no classical limit. The expectation is that finite friction induces damped oscillations on a transient time scale of the form $P_z(t)\sim {\rm e}^{-\gamma t} \cos(\Omega t)$. As long as $\Omega\neq 0$ the system dynamics is said to be coherent, otherwise it is incoherent. For vanishing friction $P_z(t)=\cos(\Delta t)$, while finite coupling leads asymptotically to a thermal equilibrium with $P_z(t\to \infty)\to 0$ (delocalized) or $P_z(t\to\infty)\neq 0$ (localized) as shown previously \cite{anders,rieger}.

To analyze the dynamical features in detail (cf.~Fig.~\ref{phasediagram}), we consider the case of zero temperature and $\omega_c\gg \Delta$, and start in Fig.~\ref{slarge} with $P_z(t)$ in the range $1/2\leq s\leq 1$ of spectral exponents. For $s$ fixed,  PIMC simulations display with increasing coupling $\alpha$ a changeover from coherent to incoherent motion, i.e., from damped oscillations to overdamped decay \cite{thorwart}.  It appears in that domain in parameter space where the thermal state is delocalized (Fig.~\ref{phasediagram}).  Accordingly,  the critical coupling strength for the coherent-incoherent turnover  $\alpha_{CI}(s)$ is always smaller than the critical coupling $\alpha_c(s)$ for the QPT.  In limiting cases one finds $\alpha_c(s=0.5)\approx \alpha_{CI}(s=0.5)$, while for an Ohmic bath ($s=1$) the known result is confirmed $\alpha_c(s=1)\approx 1 > \alpha_{CI}(s=1)\approx 0.5$.

\begin{figure}
\vspace{5.cm}
\epsfig{file=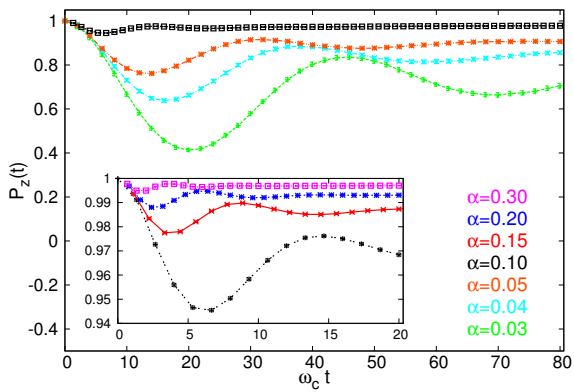,angle=-90, width=7.5cm}
\caption{Same as in Fig.~\ref{slarge} but for $s=0.25$ and values $\alpha>\alpha_c\approx 0.022$. The inset shows a blow-up for very large couplings.}\label{ssmall}
\vspace{-0.5cm}
\end{figure}

In contrast, simulations in the range $0<s< 1/2$ reveal a different scenario. Oscillatory patterns survive even for coupling strengths far beyond the critical coupling $\alpha_c(s)$ for the QPT (see Figs.~\ref{phasediagram}, \ref{ssmall}). PIMC data up to ultra-strong couplings $\alpha=30\, \alpha_c$ do not show a changeover to a classical-like decay for exponents up to $s=0.49$. The oscillation frequencies $\Omega_s(\alpha)$ in $P_z(t)$ increase with increasing coupling and exhibit a scale invariance according to $\Omega_s(\kappa\, \alpha)=\kappa\, \Omega_s(\alpha)$ (inset Fig.~\ref{ssmall} and \cite{epaps}).  Two observations are intriguing:  (i) even in the regime where friction is so strong that the thermal state resides in the classical phase \cite{hofstetter,rieger} (spin almost frozen with $P_z\sim 1\gg P_x$), does the non-equilibrium dynamics of the TSS preserve quantum coherence and (ii) the domain where these coherences survive, covers a broad range of spectral distributions up to exponents $s$ close to $s= 0.5$ (see Fig.~\ref{phasediagram}).

 To gain analytical insight, an approximate treatment is provided by the non-interacting blip approximation (NIBA) \cite{weiss}. The starting point are exact equations of motion that can be derived from (\ref{pathinte}) \cite{weiss}, i.e.,
\begin{equation}\label{eomexact}
\dot{P}_z(t)=-\int_0^t du {K}_z(t-u) \, P_z(u)\, .
\end{equation}
Within NIBA the kernel is evaluated as ${K}_{z}\approx K_{N, z}$ where
\begin{equation}\label{NIBAkern}
K_{N, z}(t)=\Delta^2 {\rm e}^{-Q'(t)} \cos[Q''(t)] \, .
\end{equation}
 Let us now analyze how incoherent decay for $P_z(t)$ may appear out of (\ref{eomexact}). In $K_{N, z}(t)$ the correlation $Q_0'(t)$ induces damping while $Q_0''(t)$ is responsible for oscillatory motion. Both functions are positive and monotonically increase in time leading to damped oscillations in $K_{N, z}(t)$. If this damping is sufficiently strong on the time scale of the bare dynamics $1/\Delta$, i.e. if the kernel is sufficiently short-ranged in time, the population $P_z(t)$ decays incoherently. We estimate this to be the case if at the first zero $t=t_*$ of the kernel where $Q_0''(t_*)=\pi/2$, the damping obeys $Q_0'(t_*)>1$. One finds this condition to be fulfilled only for $s>1/2$. There, the borderline between the two dynamical regimes is determined by that coupling strength at which $t_*=1/\Delta$, i.e.,
\begin{equation}\label{cialpha}
\alpha_{CI}(s)\approx \frac{\pi}{4 |\Gamma(s-1)| \cos(s\pi/2)}     \left(\frac{\Delta}{\omega_c}\right)^{1-s}\, .
\end{equation}
  This expression includes the known Ohmic result $\alpha_{CI}(s=1)=0.5$ and captures accurately numerical data from our PIMC. Further, it confirms our numerical finding that $\alpha_c(s)\geq \alpha_{CI}(s)$ (see Fig.~\ref{phasediagram}). For fixed $s> 0.5$, increasing friction first destroys coherent dynamics before it asymptotically induces also localization. In contrast, for spectral functions  $0< s <0.5$ a coherent-incoherent changeover never appears.

In principle, the oscillation frequency $\Omega_s$ and the damping $\gamma_s$ of  $P_z(t)$ can be extracted as complex-valued poles from the Laplace-transform of (\ref{eomexact}), i.e., $\hat{P}(\lambda)=1/[\lambda + \hat{K}_{N, z}(\lambda)]$. However, specific results can be found only for  (i) $s=0.5$ and (ii) $s\ll 1$ (for details see \cite{epaps}).

(i) In the regime  $\alpha\ll \sqrt{\Delta/\omega_c}$, the poles are given by $\lambda_{\frac{1}{2}}^\pm=\pm i\Omega_{\frac{1}{2}}-\gamma_{\frac{1}{2}}$ with  frequency
$\Omega_{\frac{1}{2}}(\alpha)\approx\Delta -\gamma_{\frac{1}{2}}$ and decay rate $\gamma_{\frac{1}{2}}(\alpha)= \alpha \Delta\, \pi\,  \sqrt{\omega_c/4\Delta}$.
There is no pole with $\lambda=0$ meaning that $P_z(t\to\infty)\to 0$ (delocalization). Apparently, the oscillation frequency {\em decreases} with growing coupling $\alpha$ while the damping rate {\em increases}. This indicates a changeover from coherent motion with $\Omega_{\frac{1}{2}}/\gamma_{\frac{1}{2}}>1$ to incoherent motion with $\Omega_{\frac{1}{2}}=0$ for larger $\alpha$.

(ii) In the domain $s\ll 1$ one may use the expansion $Q_0(t)\approx \Lambda_s\, t \exp[i\pi (1-s)/2]$ with coefficient $\Lambda_s=\int_0^\infty d\omega J_s(\omega)/(\pi \omega)= 2\alpha\omega_c |\Gamma(s-1)|$. Then, $\hat{P}(\lambda)$ has one pole at $\lambda=0$ so that $P_z(t)$ relaxes towards an asymptotic limit $P_z(t\to \infty)>0$ (localization). The other two poles are complex conjugate where the imaginary part describes oscillations in $P_z(t)$ with frequency
\begin{equation}\label{frequencylows}
\Omega_s(\alpha)\approx \Lambda_s \sin[(1-s)\pi/2] \approx \frac{2\alpha\omega_c}{s}\, .
\end{equation}
The real part corresponds to a damping rate $\gamma_s(\alpha)\approx s\, \Lambda_s$
which saturates for $s\to 0$ at $\gamma_0=2\alpha\omega_c$.  Hence, for fixed $s$, both rate and frequency {\em increase} with {\em increasing} $\alpha$ such that the dynamics of the TSS is always underdamped $\Omega_s/\gamma_s=1/s\gg 1$.
Quantum coherent dynamics persists up to arbitrarily large couplings in accordance with the numerical data (cf.~Fig.~\ref{ssmall}). Further, the above result confirms the scaling property of $\Omega_s(\alpha)$ found already in the PIMC. Note that details of the cut-off procedure may enter via $\Lambda_s$ only. Since it is determined by the low frequency behavior of $J(\omega)$, the dynamics is independent of the cut-off scheme for $\omega_c\gg \Delta$ as also seen in the PIMC (not shown). Similarly, the bare frequency of the TSS has disappeared. It only governs the dynamics for ultra-short times $t\ll 1/\omega_c$ before reservoir modes can respond. There, the bare Schr\"odinger dynamics predicts the universal quadratic time dependence $P_z(t)\approx 1- \Delta^2 t^2/2$. After this initial delocalization process, the reservoir tends to prevail with the low frequency modes in ${\cal E}$ [see (\ref{hamilton})] acting on the TSS effectively as a static energy bias of strength $-\hbar\Lambda_s$ according to the coupling $-\sigma_z {\cal E}/2$. This is in agreement with the observation that $\hbar\Lambda_s$ corresponds to the energy needed to re-organize the reservoir once the spin has flipped \cite{thoss}.

\begin{figure}
\vspace{5.5cm}
\epsfig{file=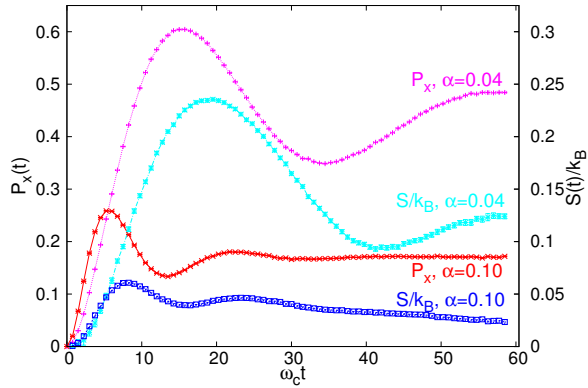, angle=-90, width=7.5cm}
\caption{Coherences $P_x(t)$ (left axis) and entropy $S(t)$ (right axis) for $s=0.25$ ($\alpha_c\approx 0.022$) at $T=0$ according to PIMC.}\label{entropy}
\vspace{-0.5cm}
\end{figure}

{\em Coherences and entanglement-- }To further verify the quantum nature of the oscillatory population dynamics, we also monitor $P_x(t)$ (Fig.~\ref{entropy}). Indeed, one observes substantial quantum nonlocality which also suggests substantial entanglement between the TSS and its surroundings. At $T=0$ this entanglement can be extracted from the von Neumann entropy $S(t)/k_{\rm B}=-w_+{\rm ln}(w_+)-w_-{\rm ln}(w_-)$ with $w_\pm=\frac{1}{2} [1\pm\sqrt{P_x(t)^2+P_y(t)^2+ P_z(t)^2}\,]$ (cf.~Fig.~\ref{entropy}). PIMC simulations reveal strong entanglement well in the regime $\alpha> \alpha_c$ (localization) on time scales where oscillatory patterns in $P_z(t)$ and $P_x(t)$ occur. For longer times and stronger coupling, entanglement decays montonically. This is in agreement with previous findings in thermal equilibrium: entanglement tends to zero in the localized phase for couplings somewhat above $\alpha_c$ \cite{hofstetter}.

\begin{figure}
\vspace{5.5cm}
\epsfig{file=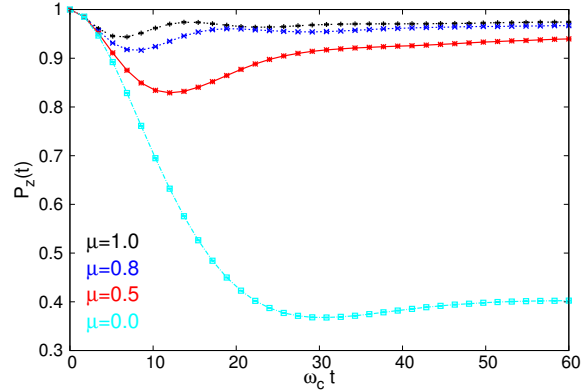, angle=-90, width=7.5cm}
\caption{$P_z(t)$ at $s=0.25, \alpha=0.1$ for various initial preparations of the reservoir with $\omega_c/\Delta=10$. See text for details.}\label{shifted}
\vspace{-0.5cm}
\end{figure}

{\em Initial preparation-- }Persistence of coherence is associated with a strong dependence on initial preparations. Of particular interest for experimental realizations are preparations, where the reservoir is out of equilibrium with respect to the initial state of the TSS. This means that in (\ref{eq:initial_prep}) one must replace $\cal{E}\to \mu {\cal{E}}$ with $\mu\neq 1$. The cloud of bath modes is shifted to the left (right) for $\mu< 1$ ($\mu>1$) with $\mu=0$ being the equilibrium orientation of the bare bath. According to Fig.~\ref{shifted}, with decreasing $\mu$ the effective oscillations frequency of the TSS decreases while at the same time the initial loss in population increases. This behavior can be understood from the fact that this preparation can equivalently be described by a time dependent bias $\epsilon_\mu(t)=(1-\mu) \dot{Q}''(t)$ of the TSS, i.e. by an additional term $\hbar\epsilon_\mu(t)\sigma_z/2$ in (\ref{hamilton}) \cite{lucke}. In particular, for reservoirs with spectral exponents $s\ll 1$ one has $\dot{Q}''(t)\approx \Lambda_s$ so that  $\epsilon_\mu\approx (1-\mu) \Lambda_s$ becomes static. Effectively, this bias adds to the bias induced by the sluggish modes in the system-reservoir coupling $-{\cal E}\sigma_z/2$ [see below (\ref{frequencylows})] to produce a net bias $\epsilon_{\rm tot}\approx -\mu \Lambda_s$. As a result the TSS regains bare dynamical features for $\mu\to 0$.
This analysis can now be extended to other preparations of the TSS alone (superpositions of $|\pm 1\rangle$) which, however, display qualitatively a similar picture. The same is true for asymmetric TSS and/or finite temperatures as long as corresponding energy scales do not exceed the bare tunneling amplitude. Results will be shown elsewhere.

{\em Summary--}We have shown that in non-equilibrium coherent dynamics can persist for ultra-strong coupling to a broadband reservoir. For a  SB model in the sub-Ohmic regime, stronger friction does not induce incoherent relaxation for spectral exponents $0<s<1/2$ even when the thermal equilibrium is almost classical.
 These findings shed light on our understanding of decoherence in open quantum systems and are thus of relevance for current experiments in nanoscale structures. The case $s=1/2$ can be realized through a charge qubit subject to electromagnetic noise \cite{ring}, while recent progress in engineering local environments for Cooper pair boxes may allow to study cases with $s<1/2$ \cite{pekola}.
 Reservoirs with $s\to 0$ have been proposed as models for $1/f$ noise in superconducting circuits where temperature enters as an effective parameter \cite{fnoise}.
 While in this latter case details need to be investigated, a promising alternative are trapped ion systems as shown in \cite{marquardt}.

\acknowledgments{We thank C. Escher, R. Bulla, A. Chin, H. Grabert and M. Grifoni for valuable discussions. Financial support from the DFG through the SFB/TRR21 and the GIF is gratefully acknowledged.}


\begin{thebibliography}{99}
\bibitem{data}Data points for the quantum phase transition are taken from Ref.~\cite{rieger}.
\end{thebibliography}
\end{document}